# The NEVOD-EAS air-shower array

M.B. Amelchakov[1], N.S. Barbashina[1], A.G. Bogdanov[1], A. Chiavassa[1,2,3],
D.M. Gromushkin[1], S.S. Khokhlov[1], V.V. Kindin[1], R.P. Kokoulin[1], K.G. Kompaniets[1],
A.Yu. Konovalova[1], V.V. Ovchinnikov[1], N.A. Pasyuk[1], A.A. Petrukhin[1],
I.A. Shulzhenko[1], V.V. Shutenko[1], I.I. Yashin[1], K.O. Yurin[1]

[1] *National Research Nuclear University MEPhI (Moscow Engineering Physics Institute), 115409 Moscow, Russia*
[2] *Dipartimento di Fisica dell' Università degli Studi di Torino, 10125 Torino, Italy*
[3] *Sezione di Torino dell' Istituto Nazionale di Fisica Nucleare – INFN, 10125 Torino, Italy*

**Abstract**

The Experimental complex NEVOD includes several different setups for studying various components of extensive air showers (EAS) in the energy range from $10^{10}$ to $10^{18}$ eV. The NEVOD-EAS array for detection of the EAS electron-photon component began its data taking in 2018. It is a distributed system of scintillation detectors installed over an area of about $10^4$ m$^2$. A distinctive feature of this array is its cluster organization with different-altitude layout of the detecting elements. The main goal of the NEVOD-EAS array is to obtain an estimation of the primary particle energy for events measured by various detectors of the Experimental complex NEVOD. This paper describes the design, operation principles and data processing of the NEVOD-EAS array. The criteria for the event selection and the accuracy of the EAS parameters reconstruction obtained on the simulated events are discussed. The results of the preliminary analysis of experimental data obtained during a half-year operation are presented.

*Keywords*: cosmic rays, primary particles, extensive air showers, scintillation detectors

## 1. Introduction

The Experimental complex (EC) NEVOD is located at the campus of the National Research Nuclear University MEPhI (Moscow, Russia) and is designed to study various components of cosmic rays on the Earth's surface. The basis of the EC NEVOD is a 2-kiloton Cherenkov water detector with a spatial array of quasi-spherical optical modules [1] which is surrounded by a large-area coordinate-tracking detector DECOR [2].

Since 2002, an experiment for studying muon component of inclined extensive air showers (EAS) using the method of local muon density spectra (LMDS) has been carried out at the Experimental Complex NEVOD. As a result of this experiment, an excess of muons in bundles in comparison with the calculations for all existing models of hadronic interactions has been discovered [3]. This phenomenon was named the "muon puzzle" [4, 5].

To calibrate the LMDS method, a facility for detection of EAS electromagnetic component (air-shower array) has been constructed around the building of the Cherenkov water detector. To detect EAS particles in this array, the scintillation counters which previously operated in the EAS-TOP [6] and KASCADE-Grande [7] experiments are used. During the array constructing, the main difficulty was the high building density of the MEPhI campus. In this regard, at the development stage the array structure has been organized by a cluster approach. According to this approach, the array consists of independent clusters installed on the roofs of the MEPhI laboratory buildings and on the ground surface between them.

## 2. NEVOD-EAS air-shower array

From 2014 to 2018 the NEVOD-EAS air shower array was constructed and commissioned inside the Experimental complex NEVOD. The main task of this array is to detect the electron-photon component of extensive air showers in the energy range from $10^{15}$ to $10^{17}$ eV. The EC NEVOD is located inside the densely built-up campus of the MEPhI University, therefore the NEVOD-EAS array detecting elements have been installed not only at the ground surface, but also on the roofs of laboratory buildings. This is the main difference between the NEVOD-EAS array and other EAS research facilities in which detecting elements are located on a plane surface.

The present configuration of the NEVOD-EAS array includes 9 clusters which are deployed around the EC NEVOD over the area of about $10^4$ m$^2$ at different altitudes. The layout of the NEVOD-EAS detecting elements is shown in Fig. 1. Clusters Nos. 1-3 and 9 are located on the roofs of laboratory buildings (height from 12 to 18 m) of the EC NEVOD, while clusters Nos. 4-8 are deployed on the ground surface. The distance between the centers of neighboring clusters is about 30 m.

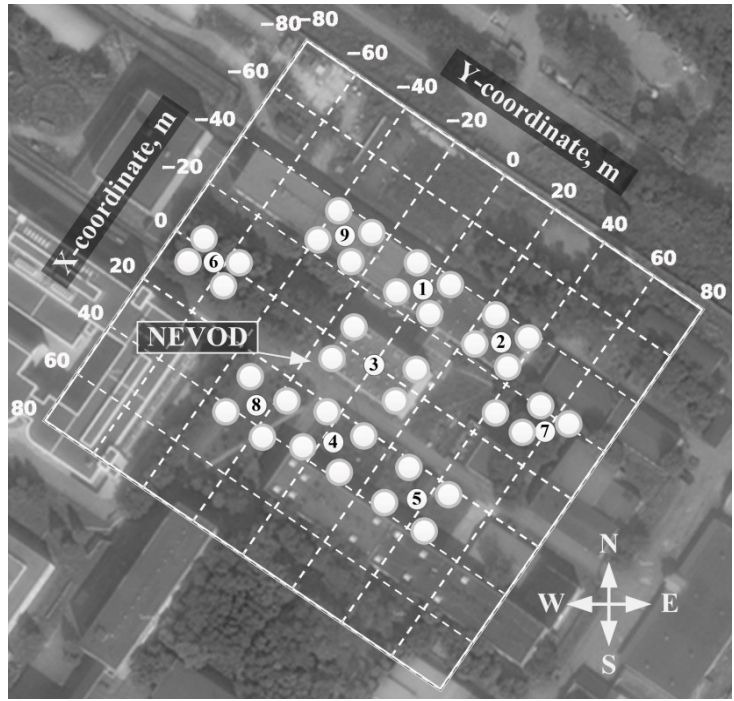

Fig. 1. Layout of the NEVOD-EAS array. The white circles are the detector stations in each of the numbered clusters.

The structure of the NEVOD-EAS array measuring system (Fig. 2) includes 2 main elements: the clusters and the central DAQ post. Each cluster represents an autonomous part of the array which provides EAS detection, selection of events according to triggering conditions (detection thresholds, coincidence multiplicity and time gate), signal digitizing, event time-stamping and data transfer to the central DAQ post of the array. In its turn, the central DAQ post ensures the synchronous operation of all clusters and their control, as well as receiving, processing and storing of experimental data from array clusters. Further, these structural elements of the array measuring system are described in details.

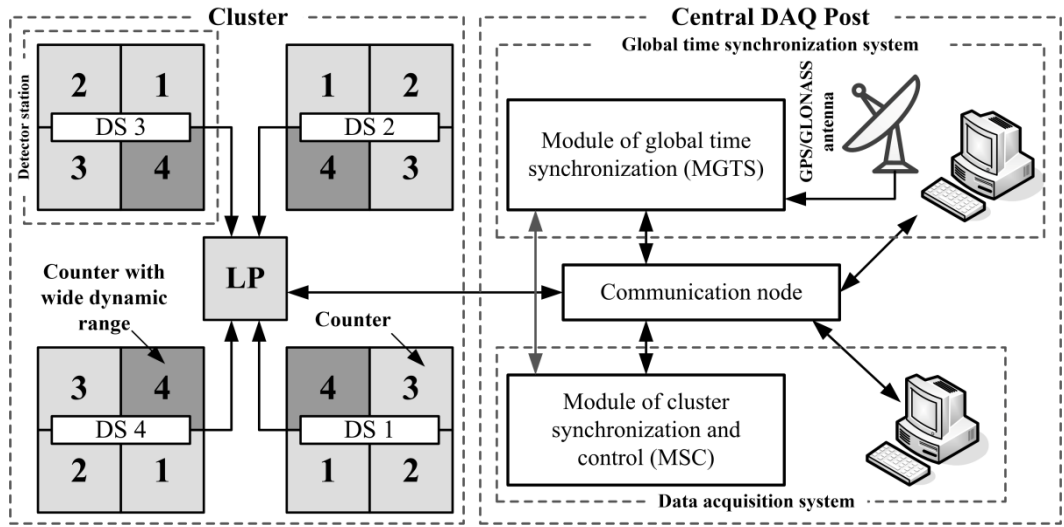

Fig. 2. Structure of the NEVOD-EAS array measuring system.

## 2.1. Cluster of the array

Cluster of the NEVOD-EAS array contains 4 scintillation detector stations (DS) combined by the local post (LP) of preliminary data processing. In a cluster, detector stations are installed at the vertices of a quadrangle (mainly a rectangle) with typical side lengths of about 15 m and form almost horizontal plane (maximal altitude difference between the cluster DSs does not exceed 1 m). Signals from DSs are fed via 25-meter long composite protected communication cables to the LP which is installed directly at the cluster location. In its turn, the LP collects and digitizes analog signals from DSs, selects events according to the triggering conditions, performs timestamping of events and transfers information of registered events to the central DAQ post.

Detector station [8] consists of 4 scintillation detectors installed inside the external protective housing (Fig. 3) and has a total area of 2.56 m$^2$ and the dynamic range of 0.3÷10$^4$ particles/m$^2$. Detectors are made by plastic scintillator NE102A with dimensions of 800×800×40 mm$^3$ and one or two photomultiplier tubes (PMT) Philips XP3462 with a 3″ hemi-spherical photocathodes mounted inside the light-isolated stainless steel pyramidal housing. Internal surface of the housing is painted with a diffusely reflecting coating that improves the light collection. These detectors were previously used in the EAS-TOP and KASCADE-Grande experiments. After their delivery to the EC NEVOD, the characteristics of scintillation detectors and their elements were studied in details. The results of these studies were published in [9].

Each of DS detectors has one "standard" PMT and only the fourth detector has an "additional" one. "Standard" PMTs are used for time and EAS particle density measurements. To achieve a dynamic range of up to 100 registered particles, the most probable response of "standard" PMTs to the passage of a single muon through the scintillator is adjusted at a value of about 13 pC. The "additional" PMT has almost 100 times lower gain than the "standard" ones (response to the passage of single muon is about 0.13 pC) and extends the DS dynamic range up to 10$^4$ particles/m$^2$.

The main element of the cluster LP is the block of electronics of the cluster of detector stations (BECDS). It digitizes analog signals from the DS PMTs, selects events according to the intra-cluster triggering conditions, timestamps experimental data and transfers them to the central DAQ post. BECDS consists of: 4-channel summator-multiplexer (SM) summing signals of 4 "standard" PMTs of each DS; two 2-channel boards of amplitude analysis (BAA) for digitizing signals from the cluster DS; the controller for managing operation of BAAs, selection of events according to the triggering conditions and transmission of information to the central

DAQ post. Controller and BAAs, based on FPGA Xilinx Spartan-3 and designed according to Euromechanics standard (19″, 6U), are installed into a specialized crate and are connected with each other via VME cross-board used to transfer control commands and BAA data. Controller and BAAs have their own clock generators with a frequency of 100 MHz which are synchronized with each other. Functional diagram of the BECDS is shown in Fig. 4.

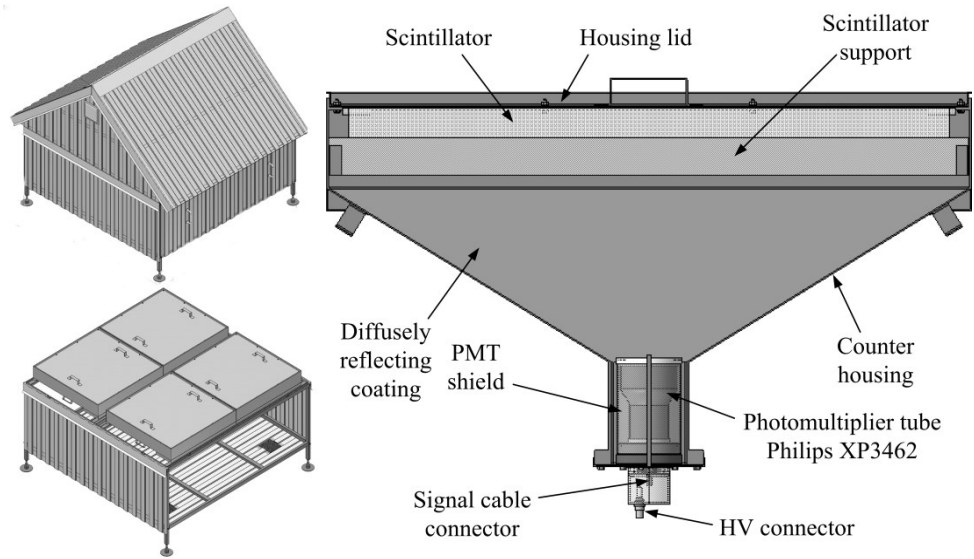

Fig. 3. The design of the NEVOD-EAS detector station.

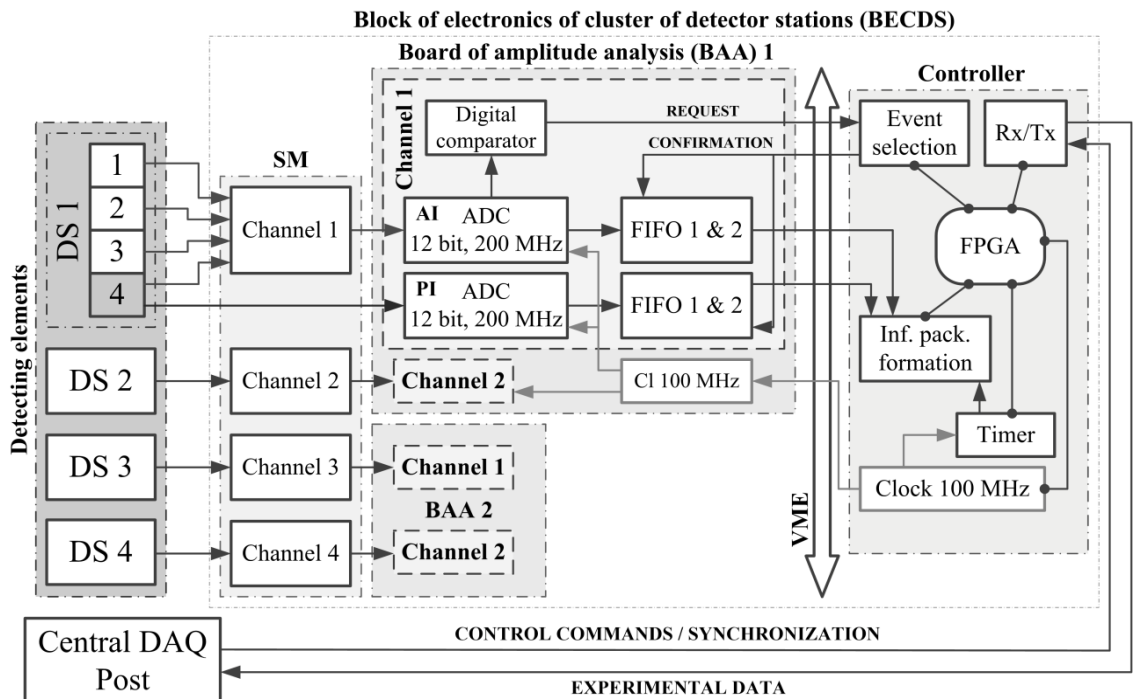

Fig. 4. Functional diagram of the BECDS of the cluster local post.

Four-channel SM [10] performs summation of signals from "standard" PMTs. Each SM channel serves the corresponding DS. The summed signals generated at the output of the summation channels are fed to the corresponding channels of the BAAs. The SM conversion coefficient (the ratio of output and input signals amplitude or charge) is 0.25.

Boards of amplitude analysis have 2 measuring channels with 2 inputs: active (AI) and passive (PI) ones. Active inputs are dedicated to readout and processing of summed signals from the corresponding DSs, while passive inputs readout signals from corresponding "additional" PMTs. At each input of the BAA channel, signals are digitized with a sampling frequency of 0.2 GHz using a 12-bit fADC. The conversion coefficients of ADC for all clusters are in range from 0.60 to 0.75 code/mV. The detection thresholds at the active inputs of BAA channels are set using 8-bit digital discriminators.

If the signal of the active input of the BAA channel exceeds a specified detection threshold, a REQUEST signal is generated and transmitted to the controller. If the number of REQUESTs received within a specified coincidence time gate exceeds a specified minimal value (multiplicity of hit DS), the controller sends a CONFIRMATION signal to all BAA channels. Upon receiving of the CONFIRMATION signal, BAAs transfer the waveforms of digitized signals stored in FIFO buffers of all channels to the controller. Then controller generates an information package, containing the signal waveforms from each input of all BAAs and the event timestamp with an accuracy of 10 ns obtained according to the internal timer of the controller. Finally, the package is transferred to the central DAQ post via fiber-optic communication line.

To store the waveforms of the digitized signals, each ADC of the BAA has 2 ring FIFO memory buffers of 1024 cells (5 μs) each. The second (backup) memory buffer is used if the first one is waiting for the CONFIRMATION signal or is transferring data to the controller. The use of two memory buffers allows to reduce the detector dead time.

Besides the BECDS, the cluster local post of preliminary data processing includes the DS high-voltage power supply, the thermoregulation system maintaining the temperature inside the LP housing in the range from 10 to 35 °C, the network equipment for receiving control commands and synchronization packets, as well as for transferring information of registered events and other service data. In outdoor conditions, all elements of the cluster LP are installed into the protective telecommunication enclosure with internal dimensions of 600×600×800 mm$^3$.

## 2.2. Central DAQ

All clusters are combined to one facility by the central DAQ post (Fig. 5) which includes two software and hardware systems for data acquisition (DAQ-system) and for global time synchronization (GTS-system), as well as the communication node ensuring network connection between all electronic equipment of the array and the storage of experimental and service data. All the central DAQ post equipment is installed into 19-inch telecommunication rack located in the control room of the Experimental complex NEVOD.

The DAQ-system transfers control commands and settings to the BECDS of the clusters, as well as receives experimental and service information from the BECDS. It consists of a computer running dedicated software and a set of modules of cluster synchronization and control (MSC) installed into the 6U crate.

The MSC module performs data exchange with the DAQ-system computer and clusters LPs, as well as synchronizes BECDS in clusters. It is based on the FPGA Xilinx Spartan-3 and operates at a clock frequency of 100 MHz. Each MSC supports parallel data exchange with 4 BECDS.

Data exchange between the MSC and BECDS is performed via fiber-optic communication lines. These communication lines are also used to transmit synchronization signals (time stamps (TS), frequency of 1 Hz) and clock frequency (CF), and thus to ensure synchronous operation of all internal clock generators and local timers in connected BECDS.

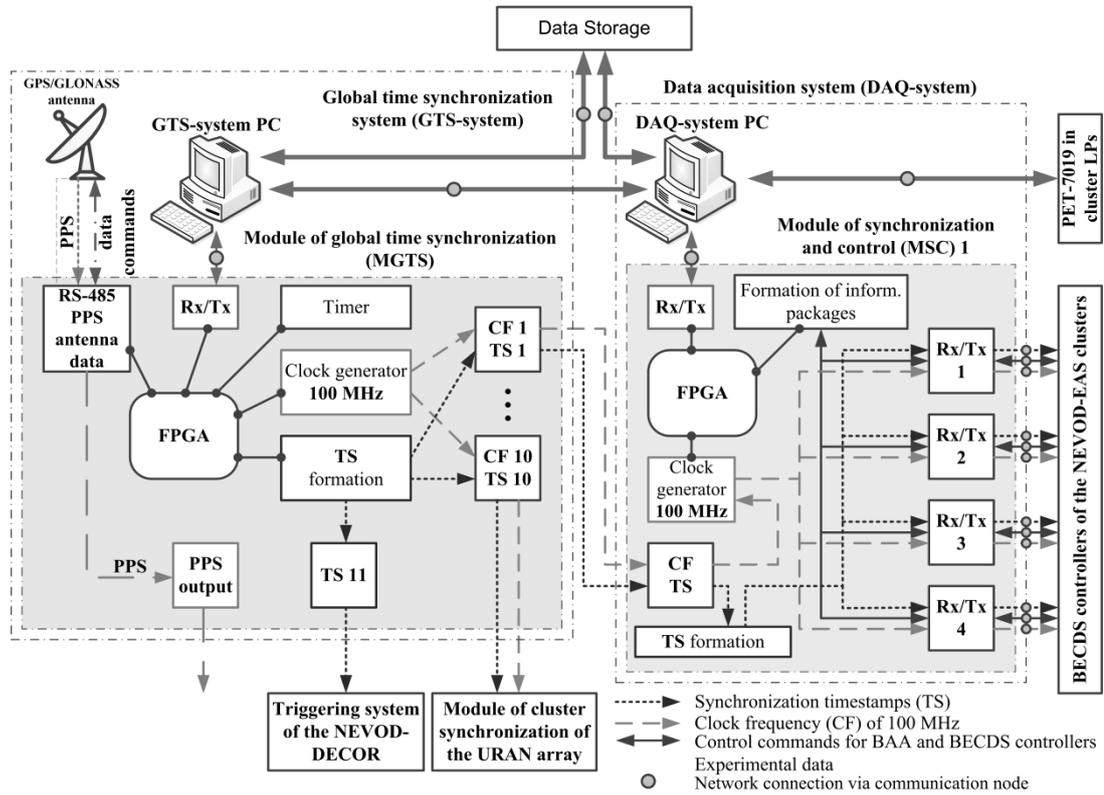

Fig. 5. Structure and functional diagram of the central DAQ post.

Data exchange between the MSC and the DAQ-system computer is performed via the Ethernet interface. Using this interface, the computer controls the MSC and the corresponding BECDS in clusters, as well as receives clusters data. In the MSC module, information packages from the clusters are supplemented with the cluster number and then transmitted to the DAQ-system computer. In its turn, the DAQ-system computer performs package parsing, analyzes data and writes them to the storage.

The GTS-system ensures synchronous operation of the array clusters and the linking of registered events to the world time. It includes the module of global time synchronization (MGTS) with connected GPS/GLONASS antenna Trimble Acutime GG and the control computer. The MGTS module is installed into the telecommunication rack along with the MSC modules, while the GPS-antenna is mounted on the roof of the Experimental complex NEVOD building.

The MGTS module is based on the FPGA Xilinx Spartan-3 and operates at a clock frequency of 100 MHz. This module provides an opportunity of independent synchronization of up to 10 devices by distributing common clock frequency (CF) and timestamps (TS). It is also equipped with 2 coaxial outputs for transmitting TS and PPS signals from the GPS-receiver for the purpose of custom device synchronization (e.g. NEVOD-DECOR triggering system [2, 11, 12]). The control of MGTS and the readout of its local time are performed by the GTS-system computer via Ethernet interface.

To ensure the synchronous operation of the MSC modules and so of the connected BECDS in clusters LPs at a common CF, this frequency is formed in one master clock generator of the MGTS module. As timers of all electronic units of the array registering system operate at the same CF, the time deviation does not occur in them, achieving a 10-ns accuracy of the array clusters synchronization. The synchronous start of local timers in all electronic units of the array registering system is performed using the MGTS timestamps which are transmitted every second with the clock frequency.

## 2.3. Data of the array

The data taking at the NEVOD-EAS array is organized in experimental series which are sequences of runs with duration of 24 hours. Each run is divided into six 4-hour intervals during which the array operates in 2 different modes: exposition (3 hours 50 minutes) and monitoring (10 minutes). In the exposition mode all array clusters detect charged particles of extensive air showers independently from each other. The intra-cluster triggering condition is at least 2-fold coincidence of DS within the time gate. For all clusters time gates with the duration of 130 ns were chosen. Detection threshold for each detector station of the cluster is about 0.75 of the most probable value of muon hump. During monitoring, the charge spectra of detector stations responses are measured in a self-triggering mode (minimal multiplicity of hit DS of each cluster is 1, registration threshold is about 0.5 of muon hump). Using these spectra the DS responses to the passage of mainly single muons are calibrated.

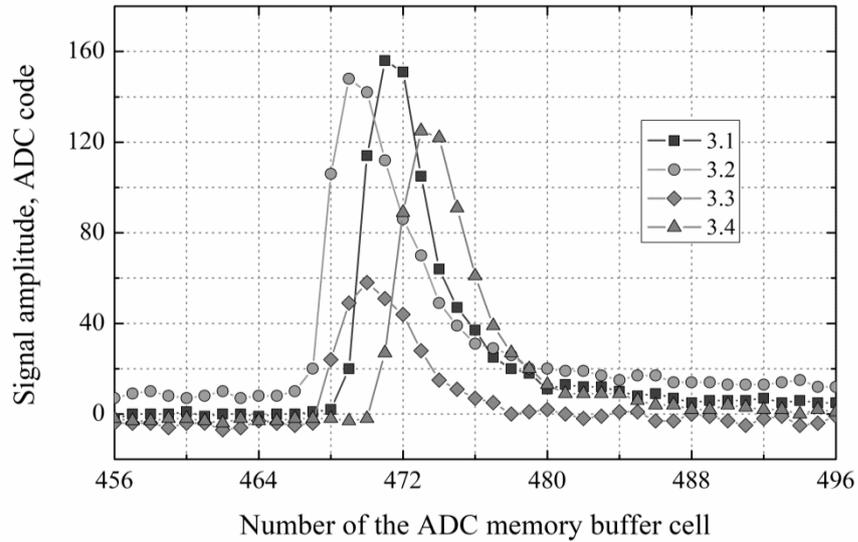

Fig. 6. Waveforms of DS signals of an event registered by cluster No. 3.

Upon detection of events, each cluster sends information packages to DAQ-system computer. The structure of information package from the cluster is the same for both exposition and monitoring modes: the order number of the cluster, 8 signal waveforms (1024 points of 5 ns each) from fADCs of BECDS channels, time of event registration according to the BECDS internal timer (timestamp). An example of digitized waveforms of signals from standard channels of detector stations in an event registered by the cluster No. 3 is shown in Fig. 6. In its turn, the computer performs parsing of information packages and then simultaneously writes the obtained data to the storage (repacked binary files) and to the database of clusters events (meta-information including unique identifier of a cluster event, cluster number, event timestamp, operation mode, path to corresponding binary file, etc).

At the end of each run, the analysis of waveforms of signals in all registered cluster events is performed. In this analysis, the mean values of the ADC baselines of BECDS channels, the amplitudes and charges of DS signals, as well as the response time of the BECDS channels (relative to the zero cell of the ADC memory buffer) are determined for each cluster event. The obtained data are added to the corresponding records in the database of cluster events. Then using the timestamps of events in separate clusters, the offline search for coincidences of clusters within the time gate with the duration 490 ns is performed. The time gate duration (490 ns) corresponds to the time required by the EAS front to pass the distance between the array detector stations most remote from each other (~ 130 m). Simultaneously, the database of array events is formed. In this database each record corresponds to the separate array event and contains unique identifier of extensive air shower, multiplicity of hit clusters and identifiers of events in triggered clusters (in database of cluster events).

The chosen duration of the time gate for the EAS events selection was validated by analyzing the hit cluster multiplicity distributions for different gate durations. Fig. 7 shows the dependences of the ratio $\beta_m(T_{gate})$ of the number of selected events $N_m$ with cluster triggering multiplicity $m$ to the number of events selected with the $T_{gate}$ = 600 ns on the time gate duration for multiplicities from 3 to 9. The value of 600 ns was chosen as the reference duration of the time gates. It is seen that all the ratios increase with the time gate duration. The values $\beta_m$ are practically constant for the time gate durations larger than 350 ns.

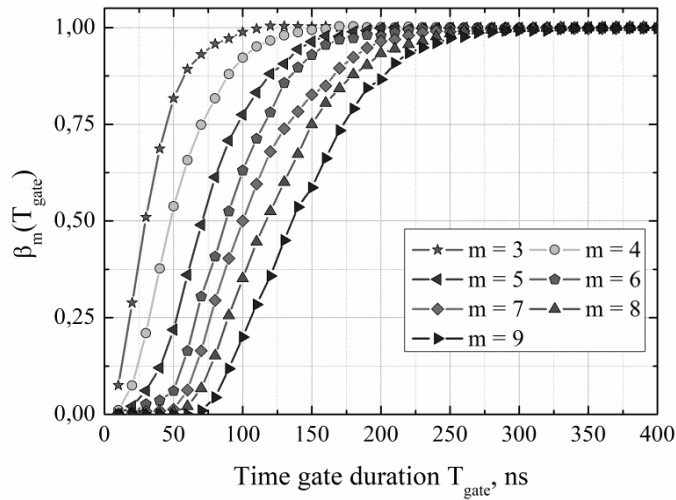

Fig. 7. Dependences $\beta_m(T_{gate})$ for different multiplicities of triggered clusters $m$ ($m$ = 3..9) in selected events.

## 3. Calibration parameters

This section describes the approaches used to obtain calibration parameters of the NEVOD-EAS air-shower array which are not only accounted for reconstruction and physical analysis of recorded events, but also for monitoring the quality of the setup operation. The calibration parameters are calculated for each run of data taking. Their analysis allowed us to determine a relatively long period of the continuous array operation from July 1 to November 10, 2020 without any critical failures. The results of the analysis of experimental data obtained during this period are presented in Section 7.

### 3.1. Delays between signals of detector stations

The EAS arrival direction is calculated using the difference of the response times of the cluster detector stations. In a particular case when a vertical EAS is registered, the detector stations of the cluster must be hit at the same time, since the shower front passes through them simultaneously. However, a non-zero differences of the response times (response delays) between the cluster DSs are observed. This is due to the difference of the PMT timing characteristics, the signal propagation time through the cables of intra-cluster communications and of the time of signal processing by the measuring channels of registering electronics. These response delays of cluster DSs must be accounted for the reconstruction of EAS arrival direction.

To obtain the relative response delays of DS in the cluster, the exposition mode data of each run are used. For all recorded events, the differences in the response times between 6 possible pairs of cluster detector stations are calculated: (1; 2), (1; 3), (1; 4), (2; 3), (2; 4) and (3; 4). As an example, the distribution of response time differences for one pair of DS (1; 2) of the cluster No. 2 is plotted in Fig. 8. The dashed curve in Fig. 8 is a Gaussian approximation. The average values of the distributions of response time differences are shifted relative to zero. These shifts are the systematic time response delays of DS to the passage of the EAS front. For the presented pair of stations, DS 2.1 on average is hit later than DS 2.2 by 3.42 ns. The width of the distribution varies slightly for different pairs of DS which is related to the distances between them. As an example, the response delays for the DSs of the cluster No. 2 are presented in Table 1.

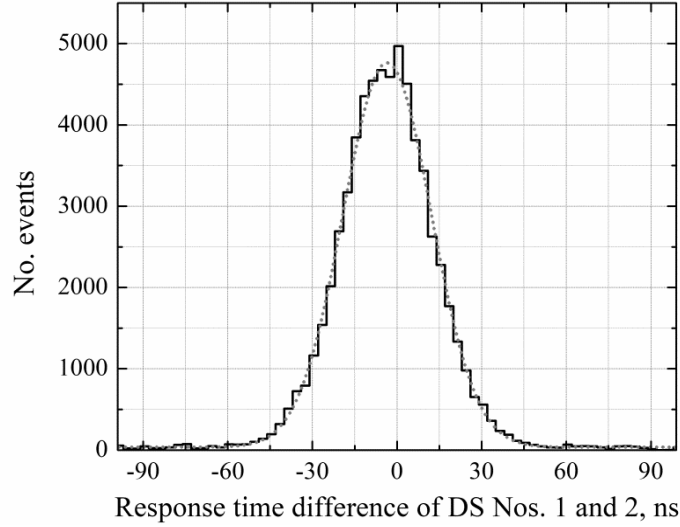

Fig. 8. Distributions of DS response time differences for DS 2.1 and DS 2.2.

Table 1. DS signal delays in the cluster No. 2

| DS pair | Delay, ns | r.m.s, ns | Distance, m |
|---|---|---|---|
| 1; 2 | -3.42 ± 0.07 | 31 | 12.4 |
| 1; 3 | 0.21 ± 0.09 | 44 | 18.2 |
| 1; 4 | -1.36 ± 0.07 | 34 | 13.3 |
| 2; 3 | 3.53 ± 0.07 | 34 | 13.3 |
| 2; 4 | 2.06 ± 0.09 | 43 | 18.2 |
| 3; 4 | -1.36 ± 0.07 | 32 | 12.4 |

### 3.2. Response to single particles

The EAS size is defined as the number of charged particles ($N_e$) in the shower at the observation level. In this work, we consider the energy deposit of particles forming muon hump as an equivalent of EAS size unit. An example of muon hump measurements for DS 1.1 is presented in Fig. 9. The position of the right peak of distribution (the so-called "muon hump") is determined by its approximation with the Gumbel function [13].

At the construction stage, the responses of array DSs were specially adjusted to ensure the positions of muon humps in the range from 10 to 20 pC. This parameter is regularly monitored during continuous data taking. Fig. 10 shows the behavior of the muon humps of detector stations of cluster No. 6 during the period selected for a further analysis. The relative deviations of muon hump positions from their mean values are of about 3 %.

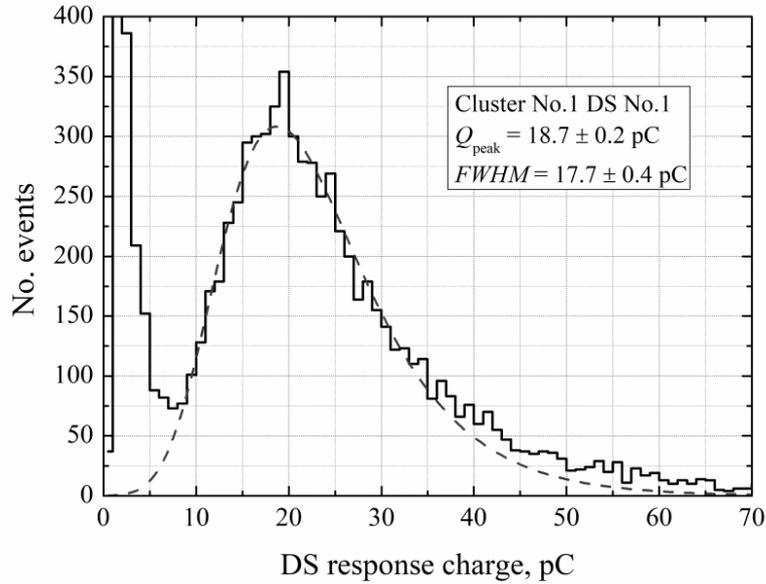

Fig. 9. Charge spectrum of responses of DS 1.1 measured in self-triggering mode. The dashed curve shows muon hump approximation by the Gumbel function.

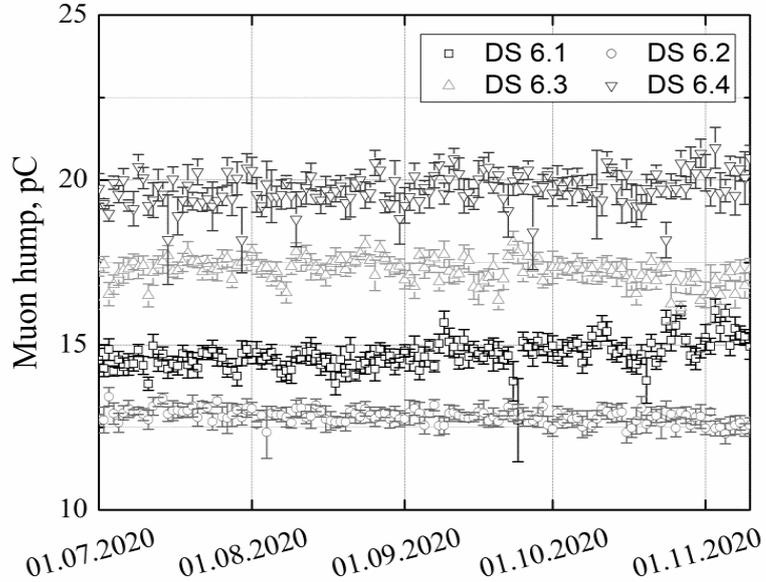

Fig. 10. Temporal behavior of muon humps positions of DSs in the cluster No. 6.

### 3.3. Cross-link coefficients for additional PMTs

A wide dynamic range of detector stations is achieved using the readout of signals from 2 measuring channels (see Section 2.1): "standard" and "additional" ones. The "standard" channel range is limited to ~ 200 particles. The "additional" channel allows us to extend it by at least 50 times. To calibrate responses of these two channels with each other, the cross-link coefficient is used.

The cross-link coefficient is calculated using the data of events recorded in the exposure mode. To perform this calibration we must define the range in which both the "standard" and the "additional" PMT have a measurable signal (the cross-link range). The lower limit of the cross-link range corresponds to the beginning of the linear transformation region of the "additional" PMT in which its registration efficiency is close to 100%. The upper limit of the cross-link range is determined by the maximal input signal for the fADC of the "standard" channel (see section 2.1). The cross-link coefficient is determined (for each single run) as the ratio ($\alpha$) of the response charges of signals at the "standard" and "additional" channels within the specified boundaries. Fig. 11 shows the α-ratio of DS 6.2 as a function of the "standard" channel response charge. The cross-link coefficient of DS 6.2 is 423 ± 38. The accuracy of cross-link coefficient calculation depends not only on the correct choice of the boundaries, but also on the number of registered events falling within the cross-link range. The counting rate of such events is about 100 per day, but, unfortunately, most of them are concentrated at the beginning of the range (see Fig. 11).

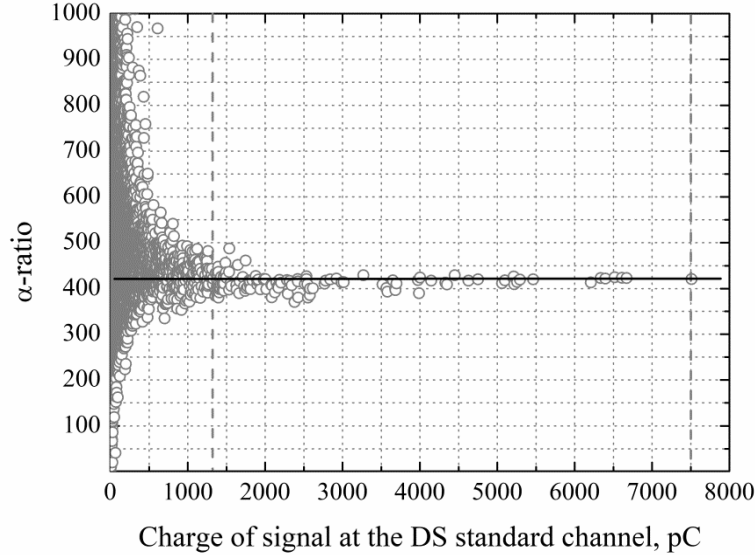

Fig. 11. The α-ratio as a function of "standard" channel response charge for DS 6.2. Dashed lines show the cross-link range boundaries. Horizontal line is the linear fit.

The values of the cross-link coefficients are practically stable over the time (Fig. 12). Therefore, in the calculations, for each DS of the array the value averaged over the entire selected period was used. The values of cross-link coefficients can be validated by the comparison of the charged particles density distributions obtained with "standard" and "additional" channels of detector stations. As shown in Fig. 13, these distributions coincide in the overlapping range of densities.

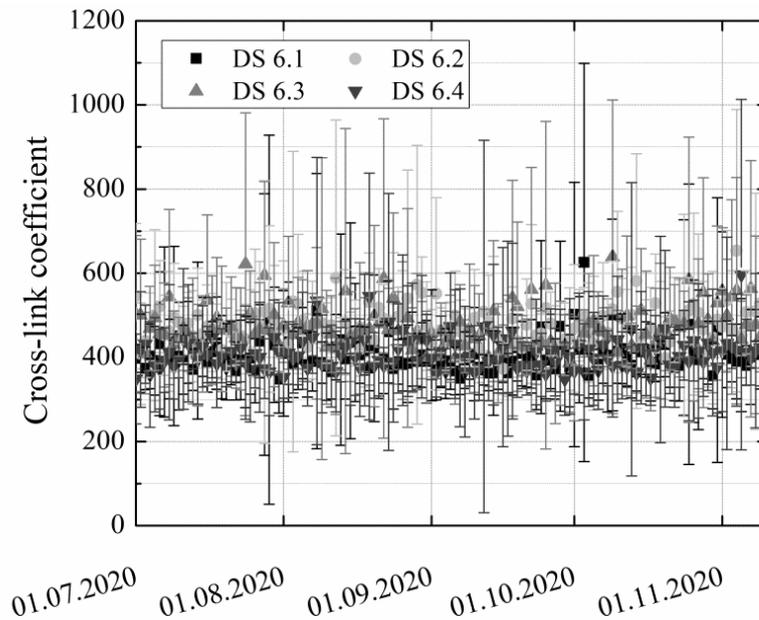

Fig. 12. Temporal behavior of cross-link coefficients of DSs in the cluster No. 6.

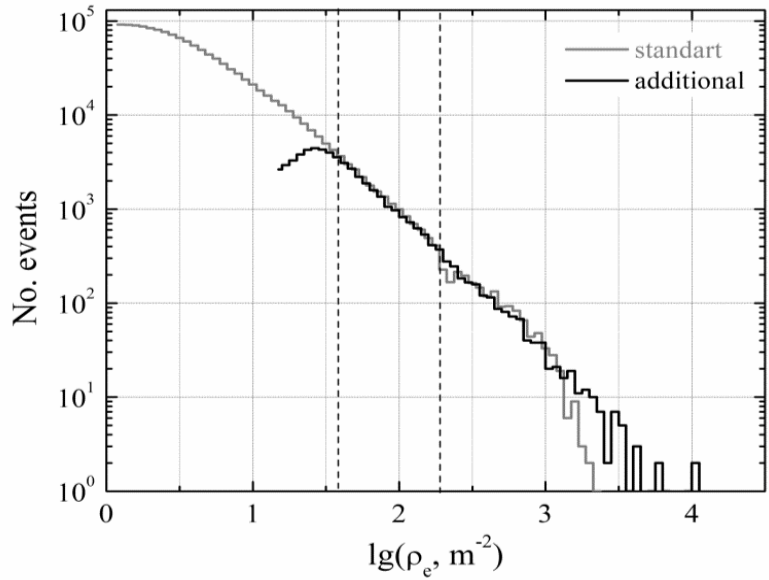

Fig. 13. Charged particles density distributions obtained with "standard" and "additional" channels of DS 6.2. The vertical dashed lines show the cross-link range boundaries.

## 4. Multiplicity of hit clusters in array events

The array event (see Section 2.3) can contain from 1 to 9 hit clusters. The dependence of the counting rate of array events, recorded during the period from July 1 to November 10, 2020, on the multiplicity of hit clusters is shown in Fig. 14. The form of the dependence is close to a power-law function. The counting rate of all array events is about one million events per day.

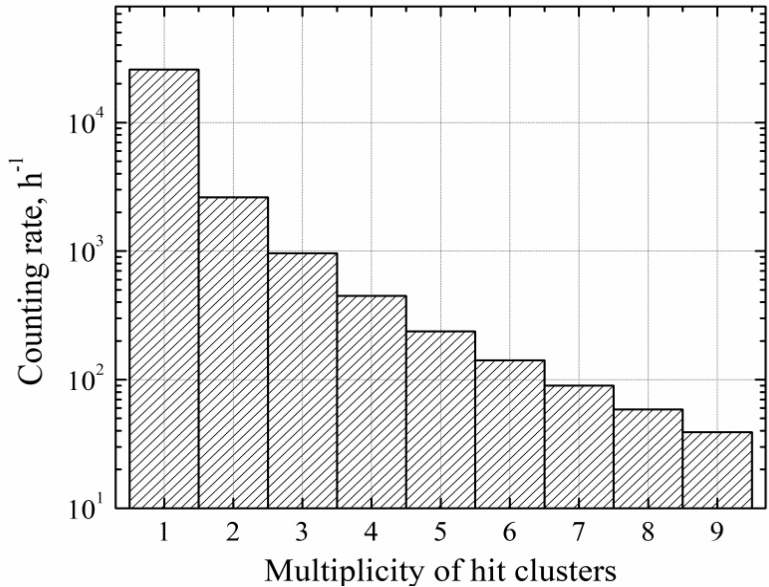

Fig. 14. Averaged counting rates of array events with different multiplicity of hit clusters.

To analyze the data on the array events counting rates, we used time intervals of about 3 hours 50 minutes between two consecutive starts of the monitoring mode. Counting rates of events with each multiplicity of hit clusters correlate with the atmospheric pressure. As an example, the daily averaged counting rate of events with 5 hit clusters is shown in Fig. 15. One can see a clear anticorrelation between the counting rate and the atmospheric pressure. Based on the obtained values of daily average counting rate, the barometric coefficients for events with different multiplicity of hit clusters were estimated (Fig. 16). The obtained dependence of barometric coefficient on the multiplicity of hit clusters indicates that the barometric effect is practically the same for the multiplicities higher than 3 and is close to -1.21 %/Torr (-0.91 %/mbar). This value is in agreement with the results of EAS registration [14] with the threshold energy of 1 PeV (-1.04±0.13 %/mbar). For multiplicities less than 4 hit clusters, the barometric effect decreases.

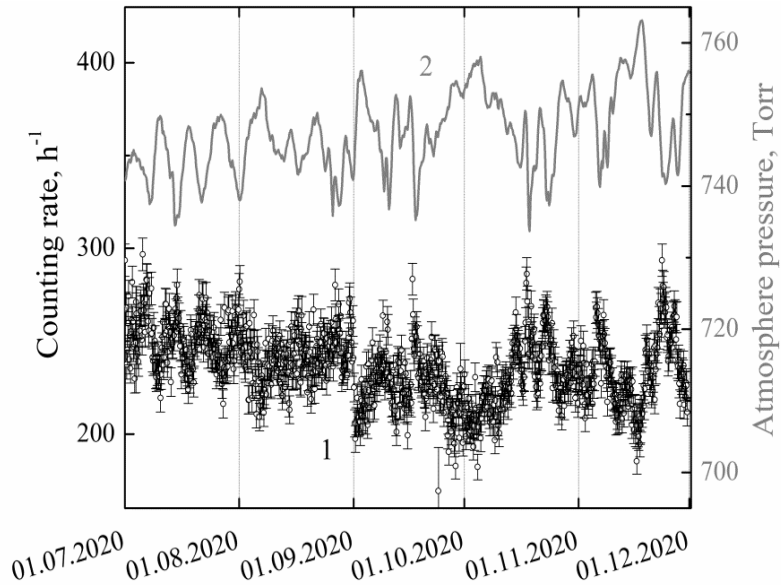

Fig. 15. Counting rate of array events with 5 hit clusters (1) and atmospheric pressure (2) during the period from July 1 to November 10, 2020.

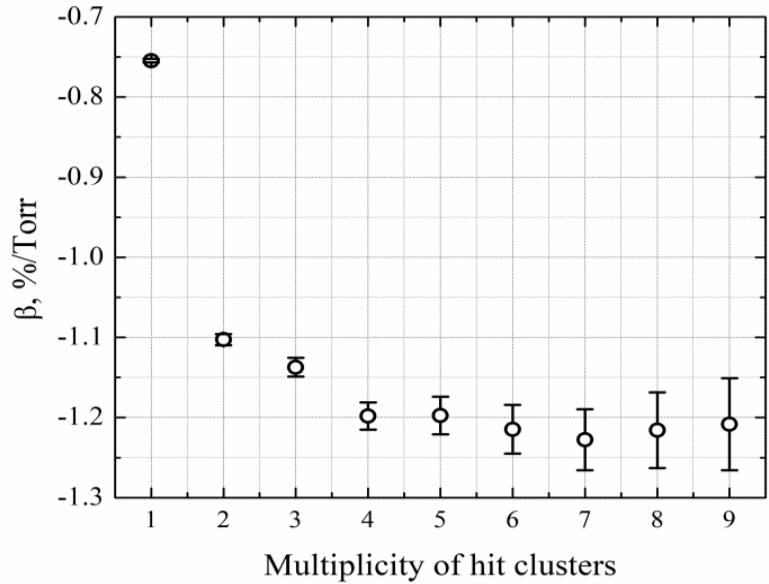

Fig. 16. Barometric coefficients for events with different multiplicity of hit clusters.

## 5. Method for EAS event reconstruction

Reconstruction of EAS event includes two independent stages. First, the EAS arrival direction is determined, then, taking into account the obtained data, all other parameters are calculated.

To reconstruct the arrival direction, the assumption of a plane front of the air-shower and the $\chi^2$ method are used. The reconstruction is based on the mutual independence of the NEVOD-EAS array clusters; local arrival directions of shower are determined by each cluster. To reconstruct local direction, at least 3-fold coincidence of DSs in cluster event is required. At this the relative response delays of DS in cluster (see section 3.1) have to be accounted. The resulting direction of EAS is the vector averaged over the local directions in clusters.

Reconstruction of other EAS parameters is performed by fitting the particle density measured by the NEVOD-EAS detector stations with the lateral distribution function [15]:

$$\rho(N_e, R, S) = \frac{N_e}{R_M^2} \cdot \left(\frac{R+\Delta}{R_M}\right)^{S-2} \cdot \left(1 + \frac{R+\Delta}{R_M}\right)^{S-4.5} \cdot \frac{\Gamma(4.5-S)}{2\pi\Gamma(S)\Gamma(4.5-2S)}, \quad (1)$$

where $N_e$ is the shower size, $R$ is the distance to the shower axis, $S$ is the shower age, $R_M$ is the Moliere radius (70.8 m), $\Delta$ is the modification term (0.8 m). The Moliere radius ($R_M$) determines the size of the cylinder containing 90 % of the particles (energy) of the shower. The modification term in formula (1) is equal to the size of scintillation detector and was introduced to avoid the uncertainty of the LDF at zero when the EAS axis hits the DS.

The EAS parameters (core location, shower age $S$ and shower size $N_e$) are determined by the iterative simplex method. At each iteration step, the value of the likelihood function is calculated. The values of parameters are considered optimal if the function reaches its minimum.

The detector stations do not directly measure the number of EAS particles that passed through them in an event. But the charge of the DS response is connected with the total energy deposit of all particles passed through the scintillators. Therefore, to reconstruct EAS parameters, it is necessary to estimate the DS response in terms of the number of particles. For this, we use the most probable response of muon hump ($Q_{hump}$) (see Section 3.2).

The detector station has two channels for measuring the response charge which are interconnected with each other by the cross-link coefficient (see Section 3.3). The maximal value of the DS response charge measured at the channel is determined by the upper limit of its dynamic range. If the DS response charge at the "standard" channel ($q_1$) exceeds its maximal value ($q_{1max}$), then the information from "additional" channel ($q_2$) is considered. Therefore, the DS response charge (Q) can be determined as follows:

$$\begin{cases} Q = q_1, & q_1 < q_{1max} \\ Q = q_2 \times K_{cl}, & q_1 > q_{1max} \\ Q = q_{2max} \times K_{cl}, & q_2 > q_{2max} \end{cases} \quad (2)$$

where $K_{cl}$ is the cross-link coefficient. The excess path length of particles within the scintillator for inclined EAS can be compensated by introducing a factor of $\cos\theta$. Thus, the number of particles $n$ detected in DS is determined by the formula:

$$n = \frac{Q\cos\theta}{Q_{hump}}. \quad (3)$$

The parameters of extensive air shower (axis position, size and age) are reconstructed by the minimization of the logarithmic likelihood function $L$:

$$L = -\ln\left(\prod_{i=1}^{k} P(n_i, N_i)\right) \quad (4)$$

where $P(n_i, N_i)$ is the probability to detect $n_i$ particles in the $i$-th DS at the expected value of $N_i$ particles, and $k$ is the number of hit DS in the array. For the current values of EAS parameters, the expected number of particles which passed through the DS with an area $s_{DS}$ is calculated by the formula:

$$N_i = \rho(N_e, R, S) s_{DS} \cos\theta. \quad (5)$$

## 6. Simulated EAS events

Simulation of the NEVOD-EAS response to the passage of air-showers includes two main stages: simulation of extensive air showers using the CORSIKA program (version 7.6400) [16] and simulation of the array response using the Geant4 program (version 10.5) [17, 18].

As primary particles we used protons. The azimuthal angle was simulated by a uniform distribution in the range from 0° to 360°. Hadronic interactions were simulated using the following models: QGSJET-II-04 [19, 20] for energies higher than 80 GeV and FLUKA 2011 [21, 22] for lower energies. The electron-photon component of simulated extensive air showers was calculated using the EGS4 [23] and NKG options included to CORSIKA program. The threshold energy (secondary particles with energies lower than the threshold one are not tracked during simulation) for hadrons and muons was 50 and 10 MeV, respectively, and 0.05 MeV for electrons and gammas. The height of the observation level was 160 m above sea level. The U.S. standard atmosphere [24] was used. The horizontal and vertical components of the magnetic induction vector of the Earth's magnetic field were 16.7 and 49.8 µT, respectively. The angle between the direction of X-axis of the coordinate system of the Experimental complex NEVOD and the direction to the North magnetic pole was 156.4°.

The response of the NEVOD-EAS array to the passage of extensive air showers was simulated using the Geant4 program. To describe particles and physical processes, one of the standard sets (PhysicsLists) QGSP_BERT was used. The geometry of the NEVOD-EAS detector stations and buildings surrounding the array was simplified. In the Geant4 model, each detector station is represented as 4 scintillators with dimensions of 800×800×40 mm$^3$ installed in the same way as in the real DS (see Fig. 3). The nearby buildings in the model have the same geometry, coordinates and composition as their real analogs, but their internal structure is ignored. The buildings are represented by the empty square-angled parallelepipeds with the wall width of 60 cm. The response of the array is the information on energy deposit of EAS particles in the DS scintillators and DS response times. The response time of detector station in the Geant4 model is determined as a time (form the start of event simulation) when the accumulated energy deposit of particles passed through the given DS exceeds the detection threshold.

The analysis of simulated events has been performed for two purposes. The first one is to obtain criteria for selection of real events. The second one is to estimate the accuracy of EAS parameters reconstruction. For this, EAS events were simulated for fixed primary particle energies and zenith angles. The core positions of extensive air showers were generated uniformly within the rectangular boundaries of the NEVOD-EAS array (see Fig. 1). For simulated events the DS measurement is calculated as the energy deposit of EAS particles crossing the scintillator. The mean energy loss of vertical minimum ionizing particles (8.2 MeV) was taken as equivalent of one particle for the analysis. The number of simulated events for various combinations of zenith angles and energies of primary protons is given in Table 2.

Table 2. Simulation statistics.

| $E_0$, PeV | No. events | |
|---|---|---|
| | $\theta = 0°$ | $\theta = 30°$ |
| 0.3 | 3000 | 3000 |
| 1 | 1000 | 1000 |
| 3 | 1000 | 1000 |
| 10 | 500 | 500 |

The accuracy of EAS parameters reconstruction depends on the location of EAS core within the boundaries of the array. It is worse when EAS core is located at the periphery of the array. At the same time, the selection of events should be done based on measured values, but not on the reconstructed ones, i.e. the responses of array detector stations or clusters. From general considerations, there is a high probability that the EAS axis is located in the central region of the array if the maximal response is observed in the station of the central cluster, as well as if the sum of responses of DSs in this cluster is greater than in the other ones, hereinafter "central events". Additional cut for arrival direction zenith angle (θ < 30°) should be implemented to exclude the effect of shadowing of detector stations, which are installed on the ground, by the surrounding buildings.

### 6.1. Detection efficiency

The trigger conditions of the NEVOD-EAS array include 2 parts: cluster ("hardware") trigger generated by the BECDS controller of the NEVOD-EAS array cluster; array ("software") trigger which is a condition for offline selection of EAS events according to the minimal multiplicity of triggered clusters. To estimate the registration efficiency of EAS under various trigger conditions and to choose the optimal ones, we have analyzed the set of simulated EAS with fixed energies and fixed zenith angles of arrival direction. All events with shower core located within the boundaries of the array were taken into account.

Fig. 17 shows the dependences of the EAS detection efficiency on the energy of the primary protons for two zenith angles of arrival direction (0° and 30°) detected under two trigger conditions differing in the "software" part: at least 5 or 7 triggered clusters. Both trigger conditions have the same "hardware" part: at least 2-fold coincidence of DS in a cluster. Here, the efficiency was defined as the ratio of the number of registered showers with multiplicity of triggered clusters not less than 5 or 7 to the number of simulated events. The shown dependences have the expected behavior: the softer the "software" part of the triggering condition is, the earlier a 100-percent efficiency of EAS registration is achieved. As seen from the figure, the showers generated by primary protons with energy higher than 1 PeV and with arrival direction zenith angles of up to 30° are registered with efficiency more than 70% if we select the trigger condition with at least 7-fold coincidence of clusters in an array event.

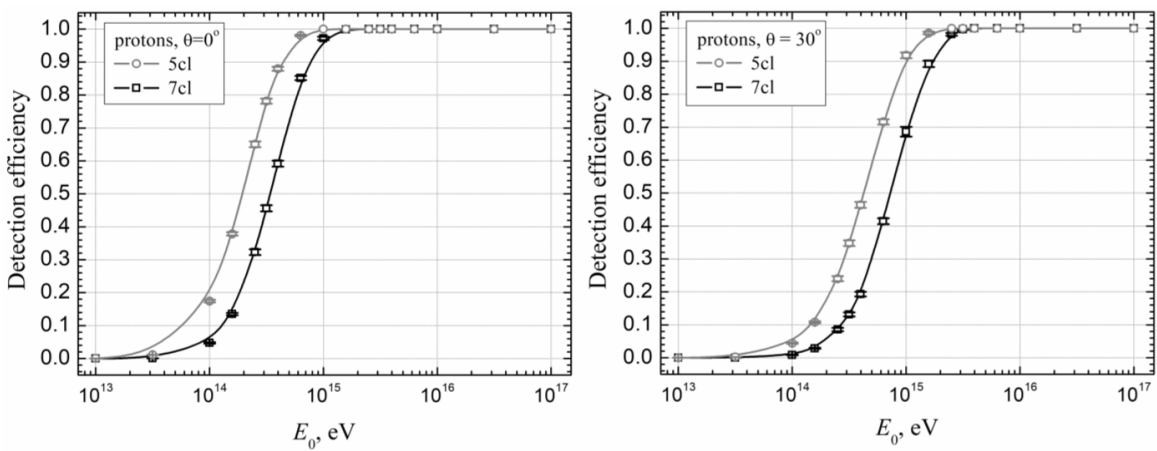

Fig. 17. Detection efficiency as a function of primary energy for fixed arrival direction zenith angles (0° and 30°) and two trigger conditions: at least 5 or 7 triggered clusters in an array event.

### 6.2. Arrival direction reconstruction

Generally, the arrival direction of extensive air shower is reconstructed using the time of flight method. We reconstruct EAS arrival direction by the data of individual clusters. The information on hit time of detector station in cluster is discarded if its hit delay (relative to the first hit DS) exceeds 130 ns.

To estimate the accuracy of the arrival direction reconstruction, we have analyzed the distribution of events by the deviation angle between the simulated and reconstructed EAS directions. The accuracy can be defined from the distribution by 68% of the events having deviations less than it. Accounting that the NEVOD-EAS array area is not large, the accuracy of direction reconstruction depends on the type of events selected for the analysis. For energies greater than 1 PeV, the accuracy is from ~ 5° for all detected EAS to ~ 1° for central events.

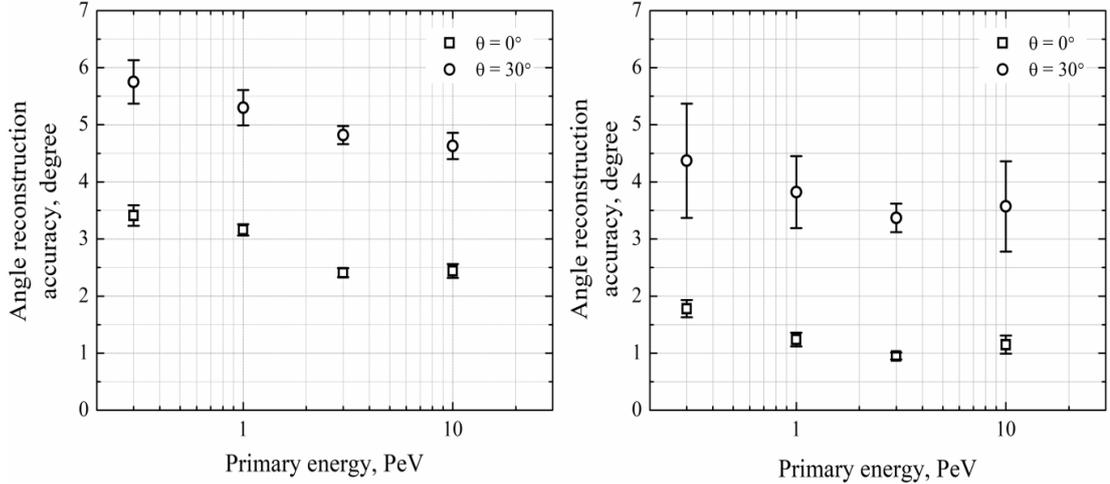

Fig. 18. Accuracy of arrival direction reconstruction as a function of primary energy for all simulated events (left) and for central events only (right).

### 6.3. Core position and size reconstruction

The estimation of the reconstruction accuracy of EAS core position and size was carried out using simulated events with fixed primary particle energies and zenith angles of arrival direction. We have analyzed both the entire set (see Table 2) of simulated events and the events falling in the central region of NEVOD-EAS array (central events) which were selected according to the criteria described at the beginning of this section.

The accuracy of core position reconstruction can be obtained from the distribution of the distances between the simulated and reconstructed points. The obtained accuracies of EAS core position reconstruction for different energies and zenith angles are presented in Fig. 19. It is clearly seen that core position reconstruction accuracy is several times better for the central events.

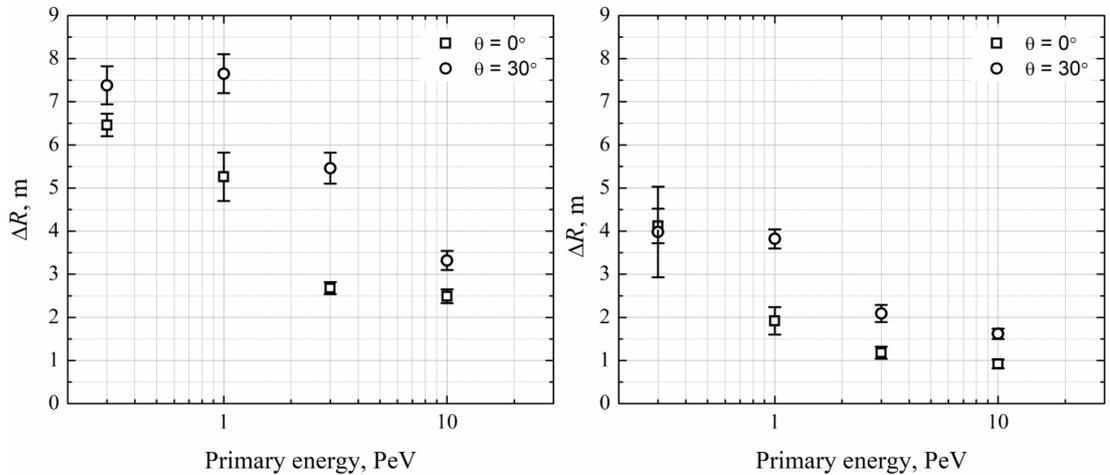

Fig. 19. Accuracy of EAS core position reconstruction as a function of primary energy for all simulated events (left) and only central events (right).

A similar situation is observed when the EAS size is reconstructed. For some peripheral events, the differences between simulated and reconstructed size values can reach several orders of magnitude. Fig. 20 shows the dependence of the relative EAS size reconstruction accuracy ($\Delta N_e/N_e$) on the logarithm of reconstructed size ($\lg N_e$) for the central events.

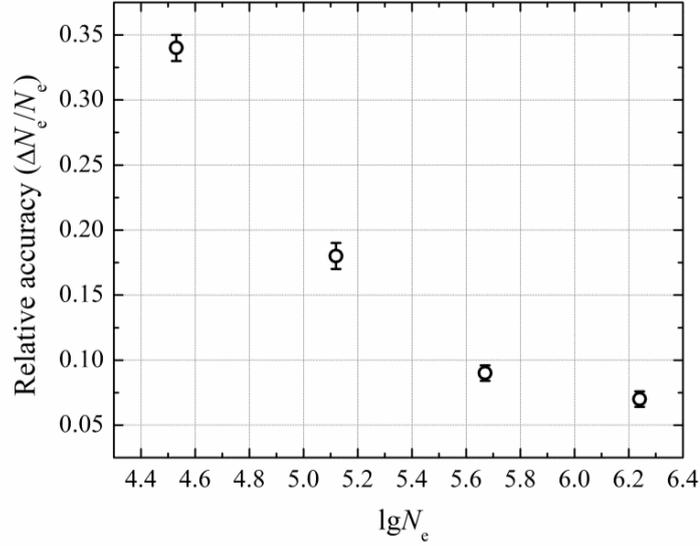

Fig. 20. Relative accuracy of EAS size reconstruction as a function of size logarithm.

## 7. First experimental data

For the analysis of the NEVOD-EAS array data, we have selected a period of relatively stable operation from July 1 to November 10, 2020. The total live time of the array during this period is about 3 000 hours. The average counting rate of the array is 728 thousand events per day. Only array events having at least 7 hit clusters and satisfying the criteria of location of the core in the NEVOD-EAS central region (see Section 6) were analyzed. The number of events selected for the analysis was 72158.

The distribution of the reconstructed core positions in the plane of the central cluster location is shown in Fig. 21. A specific geometric shape of this distribution is conditioned by the competition of neighboring clusters for the maximum response in the event.

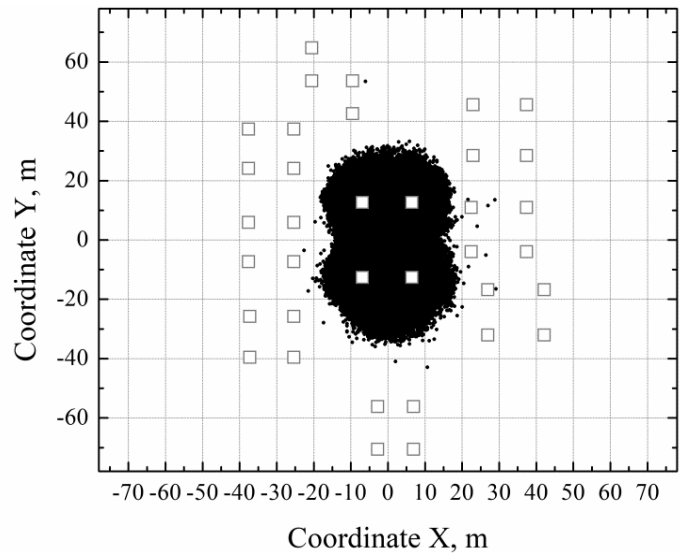

Fig. 21. Distribution of the core positions of reconstructed EAS in the plane of the central cluster location. The squares show locations of detector stations.

The distribution of air-showers by the zenith angle of arrival direction has a form shown in the Fig. 22 and somewhat differs from the expected one obtained at other EAS arrays located at the sea level [15]. Such inconsistency can be explained by the shadowing effects of the buildings and different altitudes of clusters location and non-plane structure of the array. This effect is clearly seen in the distributions of EAS by the azimuthal angle of arrival direction (Fig. 23). The azimuthal angle distribution is quite uniform for zenith angles from 5° to 10°. For larger zenith angles, from 10° to 20°, the DS shielding appears and the distribution becomes non-uniform. For zenith angles larger than 20° the geometric factor of the array clearly reveals itself (see Fig. 1). For the same primary energy, the probability of triggering more than 6 clusters is greater when air-shower moves along the Y-axis than if it arrives along the X-axis.

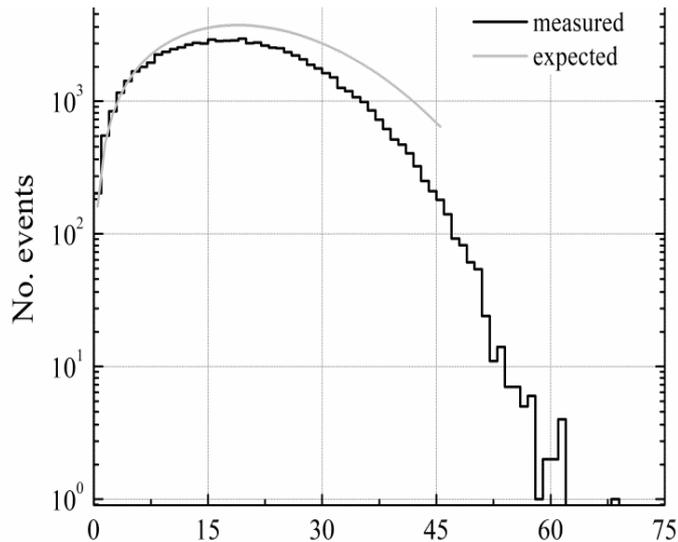

Fig. 22. Distribution of central EAS events by zenith angles of arrival direction.

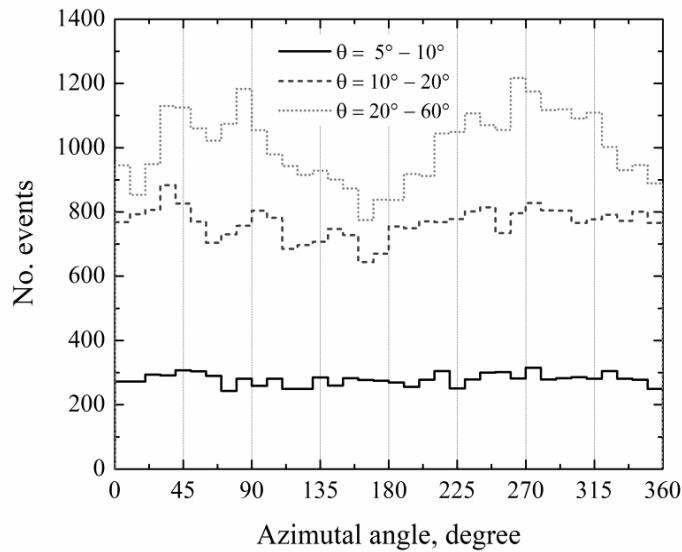

Fig. 23. Distributions of central EAS events by azimuthal angles of arrival direction in three ranges of zenith angles (5° – 10°, 10° – 20° and 20° – 60°).

The EAS size spectrum obtained at the NEVOD-EAS array as a result of analysis of central events arriving at zenith angles from 0° to 30° is shown in Fig. 24. The spectrum has a power-law form. The slope index of the spectrum obtained by its approximation in the range of $\lg N_e$ from 5.5 to 6.5 (indicated in Fig. 24 by the straight vertical lines) is -2.49±0.03. This obtained slope index value is consistent with the results of other air-shower experiments [15].

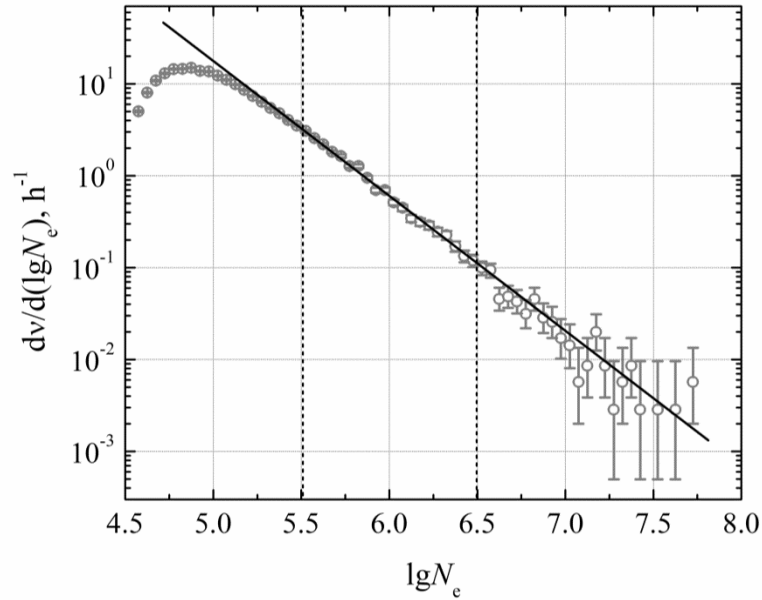

Fig. 24. Differential EAS size spectrum obtained at the NEVOD-EAS array.

## Conclusion

The NEVOD-EAS air-shower array is a part of the Experimental complex NEVOD. As in many other air-shower arrays, its registering system is organized by a cluster principle. But the main difference between the NEVOD-EAS array and other EAS research facilities is that its clusters are located at different altitudes: on the building roofs and on the ground surface.

The array covers the energy range from $10^{15}$ to $10^{17}$ eV. At such energies of primary particles for central events, the accuracy of direction reconstruction is about 1° for vertical EAS and about 3.5° for air-showers arriving at zenith angle of 30°, the core position accuracy is better than 4 m, and the relative size reconstruction accuracy is better than 20%.

The array will be used for investigation of "the knee" and other features in the cosmic ray energy spectrum, and for the calibration of the local muon density spectrum method.

## Acknowledgements

The work was performed at the Unique Scientific Facility "Experimental complex NEVOD" with the financial support provided by the Russian Ministry of Science and Higher Education, project "Fundamental problems of cosmic rays and dark matter", No. 0723-2020-0040.

The authors would also like to express gratitude to the KASCADE-Grande collaboration and the Institut für Astroteilchenphysik of the KIT for their material contribution to the maintenance and development of the NEVOD-EAS air shower array.

The authors sincerely thank the former members of the Experimental complex NEVOD staff, namely O.I. Likiy, V.Yu. Kutovoy, N.N. Kamlev, D.S. Brovtsev, P.V. Semov, N.V. Ampilogov, N.E. Fomin, for their contribution at different stages of the array development, construction and maintenance.